\documentstyle[epsfig]{aipproc}
\begin{document}

\title{Matter-induced hadronic processes \thanks{Talk presented by W.
Florkowski.} $^,$ 
\thanks{Research supported by PRAXIS 
grants XXI/BCC/429/94 and PRAXIS/P/FIS/12247/1998, and by
the Polish State Committee for Scientific Research grant 2P03B-080-12.}}

\author{W. Broniowski$^*$, W. Florkowski$^*$, and B. Hiller$^{\dagger}$}
\address{$^*$H. Niewodnicza\'nski Institute of Nuclear Physics \\
ul. Radzikowskiego 152, 31342 Krak\'ow, Poland \\
$^{\dagger}$Center for Theoretical Physics, University of Coimbra \\
P-3004 516 Coimbra, Portugal}

%\lefthead{LEFT head}
%\righthead{RIGHT head}
\maketitle

\begin{abstract}
Two examples of ``exotic'' phenomena which become possible and
important in the presence of nuclear matter are discussed: $\omega
\to \pi \pi $ decay, and $\rho-\omega$ mixing. Significance of
these processes for the low-mass dilepton production in
relativistic heavy-ion collisions is indicated.
\end{abstract}

\section*{Introduction}
An interesting factor brought in by the presence of the medium is that
processes which are forbidden in the vacuum by symmetry principles now
become possible. The constraints of Lorentz-invariance,
$G$-parity, or isospin invariance, are no longer effective. Below we
discuss two examples of such processes: $\omega \to \pi \pi $ decay,
and $\rho-\omega$ mixing. Strictly speaking, $\rho$ and $\omega$ mix
in the vacuum but this is a negligible effect caused by the small
explicit breaking of the isospin symmetry. Similarly, the partial
width for the decay $\omega \to \pi \pi $ is only $\sim 0.2$MeV 
in the vacuum, which is again a tiny isospin-violation effect. 

In this paper we summarize the results of Refs. \cite{bfh,bfh1,rhoom}
which extend the work presented in  Refs. \cite{wolf} and \cite{dutt}.
We emphasize  that the {\it matter-induced width} for the 
$\omega \to \pi \pi $ decay is large: for $\omega$ moving with
respect to the medium with a momentum above $\sim 200\mathrm{MeV}$ the
corresponding width, at the nuclear saturation density, is of the
order of 100MeV.  We also show that even a moderate excess of neutrons
over protons in nuclear matter, such as in ${}^{208}{\rm Pb}$, can
lead to large $\rho$-$\omega$ mixing.

The in-medium broadening of the $\omega$ meson, as well as the shifts
of the positions of the resonances (due to their mixing) are examples
of the so-called in-medium modifications of hadron properties, which
are predicted in a variety of theoretical calculations
\cite{brscale,celenza,hatlee,jean,cassing,li,hatsuda,rapp,pirner,klingl,leupold,eletsky,friman2,bratko}. The
recent interest in studying such modifications (for a review see
\cite{tsukuba}) has been trigerred by the experimental
observation of the enhanced production of low-mass dileptons in
relativistic heavy-ion collisions \cite{ceres,helios}. The data are
most easily described by the assumption that either the masses of
vector mesons decrease in medium or their widths become larger.

The non-vanishing amplitude for the decay $\omega \rightarrow \pi \pi$
indicates that the processes of pion annihilation into dilepton pairs
in the $\omega$ channel are also possible, as first pointed in
Ref. \cite{wolf}.  However, due to the smallness of the $\omega
\gamma$ coupling they cannot compete with the annihilation occuring in
the $\rho$ channel \cite{bfh1,wolf}.  Nevertheless, the large width of
the $\omega$ mesons should cause a depletion in their population. In our
opinion such an effect  should be included in simulations of
heavy-ion collisions. In fact, the results of some recent transport
calculations \cite{vkoch1,vkoch2} show an excess in the dilepton yield at
$q^2=m^2_\omega$, attributed to the direct $\omega \rightarrow e^+
e^-$ decay. With an increased hadronic width of the $\omega$ a better
agreement with the data may be achieved.

The effects of the isospin asymmetry in nuclear matter for the $\rho-
\omega$ mixing were studied in Ref. \cite{dutt} in the framework of the 
Walecka model. The results of Ref. \cite{dutt} indicate that at
asymmetries such as in $^{208}$Pb and at nuclear saturation density,
the $\rho$ and $\omega$ mix with an angle of about $\sim 2\%$.
In our approach, performed on a broader footing, we show that the
matter-induced $\rho-\omega$ mixing can be in fact much larger. We
expect that it may show up, among other medium-induced effects, in 
future high-accuracy relativistic heavy-ion collisions.

\section*{$\omega \rightarrow \pi \pi$ decay in nuclear matter}

Our calculation of the $\omega \rightarrow \pi \pi$ width is done
in the framework of an effective hadronic theory. Mesons ($\omega,
\sigma,\pi$) interact with nucleons and $\Delta(1232)$. We work to the
leading order in nuclear density, hence only the diagrams shown 
in Fig. 1 are taken into account. The ``bubble'' diagram (a)
was studied by Wolf, Friman, and Soyeur \cite{wolf}, who pointed
out the significance of the $\omega-\sigma$ mixing for the
in-medium  $\omega \rightarrow \pi \pi$ decay. The ``triangle''
diagrams (equally important in any formal scheme) were taken
into consideration in Ref. \cite{bfh}. The complete set of
diagrams (a-d) was included in Ref. \cite{bfh1}.

\begin{figure}[b!] % fig 1
\centerline{\epsfig{file=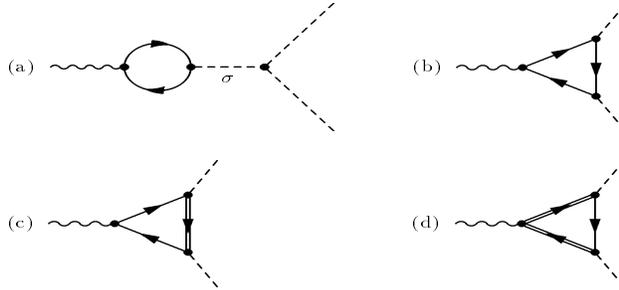,height=6.0in,width=6.0in}}
\vspace{-6cm}
\caption{ Diagrams contributing to the $\omega \to \pi \pi $ 
amplitude in nuclear medium. The incoming $\omega $ has momentum $q$
and polarization $\epsilon $. The outgoing pions have momenta $p$ and $q-p$. 
Diagrams (b-d) have corresponding crossed diagrams, not displayed.}
\label{flo1}
\end{figure}

The solid line in Fig. 1 denotes the in-medium nucleon propagator \cite{chin}
\begin{equation}
G(k) = (\not \hspace{-0.08cm}{k}+M)\left[\frac
1{k^2-M^2+i\varepsilon }+\frac{i\pi }{E_k}\delta (k_0-E_k)\theta
(k_F-|\mathbf{k}|)\right], \label{Nprop}
\end{equation}
where $k$ is the nucleon four-momentum, $M$ denotes the nucleon mass, $E_k=%
\sqrt{M^2+\mathbf{k}^2}$, and $k_F$ is the Fermi momentum.  
Diagram (a) involves the intermediate $\sigma $-meson propagator, 
which we take in the form 
\begin{equation}
G_\sigma (k)={\frac 1{k^2-m_\sigma ^2+i\,m_\sigma \Gamma _\sigma -{\frac 14}%
\Gamma _\sigma ^2}}.  \label{sprop}
\end{equation}
Here the mass and the width of the $\sigma $ meson are chosen in such a way
that they reproduce effectively the experimental $\pi \pi $ scattering
length at $q^2=m_\omega ^2=(780\mathrm{MeV})^2$, which is the relevant
kinematic point for the process at hand. From this fit we find $m_\sigma
=789 $MeV and $\Gamma _\sigma =237$MeV. Note that $m_\omega $ and $m_\sigma $
are very close to each other, which enhances the amplitude obtained from
diagram (a) \cite{wolf}.

The double line in diagrams (c-d) denotes the $\Delta $ propagator 
\begin{equation}
G_\Delta ^{\alpha \beta }(k)=
{\frac{\not \hspace{-0.08cm} {k}+M_\Delta }{k^2-M_\Delta
^2+i\,M_\Delta \Gamma _\Delta -{\frac 14}\Gamma _\Delta^2}} 
\left[ -g^{\alpha
\beta }+{\frac 13}\gamma ^\alpha \gamma ^\beta +{\frac{2k^\alpha k^\beta }{%
3M_\Delta ^2}}+{\frac{\gamma ^\alpha k^\beta -\gamma ^\beta k^\alpha }{%
3M_\Delta }}\right].  \label{Dprop}
\end{equation}
This formula corresponds to the usual Rarita-Schwinger definition \cite
{rarita,mukho} with the denominator modified in order to account for the
finite width of the $\Delta $ resonance, $\Gamma _\Delta =120$MeV.

We assume that the $\omega NN$ and $\omega \Delta \Delta $ vertices have the
form which follows from the minimum-substitution prescription and
vector-meson dominance applied to the nucleon and the Rarita-Schwinger
\cite{rarita} Lagrangians:
\begin{eqnarray}
&&V_{\omega NN}^\mu =g_\omega \gamma ^\mu ,\\
&&V_{\omega \Delta
\Delta }^{\mu \alpha \beta }=g_\omega \left[ -\gamma ^\mu g^{\alpha \beta
}+g^{\alpha \mu }\gamma ^\beta +g^{\beta \mu }\gamma ^\alpha +\gamma ^\alpha
\gamma ^\mu \gamma ^\beta \right].  \label{omv}
\end{eqnarray}
 The results presented below do not qualitatively depend on
the form of the coupling, as long as it remains strong. The coupling
constant $g_\omega $ can be estimated from the vector dominance model. We
use $g_\omega =9$. For the $\pi NN$
vertex we use the pseudoscalar coupling, with the coupling constant $g_{\pi
NN}=$ $12.7$. The same value is used for $g_{\sigma NN}$. The $\sigma \pi
\pi $ coupling constant is taken to be equal to $g_{\sigma \pi \pi
}=12.8\,m_\pi $, where $m_\pi =139.6\mathrm{MeV}$ is the physical pion mass
(this value follows from the fit done to $\pi \pi $ scattering phase shifts
done in Ref. \cite{wolf}). The $\pi N\Delta $ vertex has the form
$V_{\pi N\Delta }^\mu =({f_{\pi N\Delta }}/{m_\pi})p^\mu {\vec T}$,
where $p^\mu $ is the pion momentum, $\vec T$ is the ${\frac 12}\rightarrow {%
\frac 32}$ isospin transition matrix, and the coupling constant $%
f_{\pi N\Delta }=2.1$ \cite{durso}.

The amplitude, evaluated according to the diagrams depicted in Fig. 1 (a-d)
can be uniquely decomposed in the following Lorentz-invariant way: 
\begin{equation}
{\cal M}=\epsilon^\mu (A p_\mu +B u_\mu +C q_\mu ),  \label{decomp}
\end{equation}
where $p$ is the four-momentum of one of the pions, $q$ is the four-momentum
of the $\omega $ meson, $u$ is the four-velocity of nuclear matter, and $%
\epsilon $ specifies the polarization of $\omega $. Our calculation is
performed in the rest frame of nuclear matter, where $u=(1,0,0,0)$. In this
reference frame the amplitude $\cal{M}$ vanishes for vanishing
3-momentum $\mathbf{q}$, as requested by rotational invariance. Hence, the
process $\omega \rightarrow \pi \pi $ occurs only when the $\omega $ moves
with respect to the medium.

\begin{figure}[b!] % fig 2
\centerline{\epsfig{file=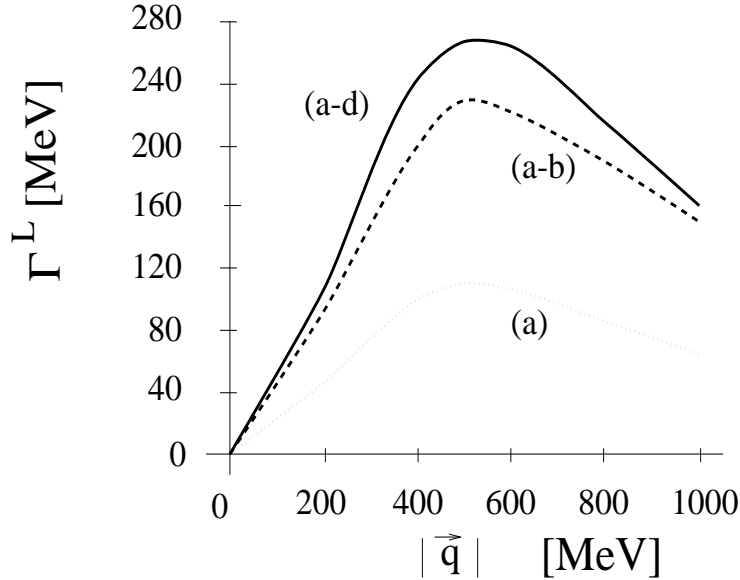,height=3.0in,width=3.8in}}
%\vspace{-5cm}
\caption{The in-medium width pf the $\omega$ meson plotted
as a function of its 3-momentum $|{\bf q}|$. The labels (a),
(a-b) and (a-d) refer to Fig.1. They indicate the diagrams
included in the calculation.}
\label{flo2}
\end{figure}

In Fig. 2 we present our numerical results at the nuclear saturation
density, $\rho _B=\rho_0=0.17\mathrm{fm}^{-3}$. We show the width
of longitudinally polarized $\omega$ mesons, $\Gamma ^L$,  as
a function of $|\mathbf{q}|$. In our calculation, we reduce the value
of the in-medium nucleon mass to 70\% of its vacuum value,
$M^{*}=0.7M,$ which is a typical number at the nuclear saturation
density. We also reduce by the same factor the mass of the $\Delta $,
\emph{i.e.} $M_\Delta ^{*}=0.7M_\Delta $, since it is expected to
behave similarly to the nucleon.  The labels indicate which diagrams
of Fig. 1 have been included. The complete result corresponds to the
case (a-d). The case (a) reproduces the result of Ref. \cite{wolf}.
The width of the transverse modes ($\sim 1$ MeV) is negligible and we
do not show it in the plot.

In Fig. 1 the mass of $\omega$ is kept at the vacuum value. For
$m^*_\omega=0.70 m_\omega$ our results decrease by a factor of 
$2.5$. Still, the widths remain substantial and the effects 
discussed above are important.

\section*{$\rho-\omega$ mixing in asymmetric nuclear matter}

For vanishing 3-momentum ${\bf q}$, the vector-meson correlator in the 
coupled $\rho^0$ and $\omega$ channels has the following structure
\begin{equation}
\Pi^{\alpha \beta }(\nu,{\bf q}=0)=\left(
\begin{array}{cc}
\Pi_\rho ^{\alpha \beta }(\nu) & \Pi_{\rho \omega }^{\alpha \beta}(\nu) \\
\Pi_{\rho \omega}^{\alpha \beta}(\nu) & \Pi_\omega ^{\alpha \beta }(\nu)
\end{array}
\right),
\label{eq:pi}
\end{equation}
where $\nu$ is the energy variable. For the diagonal parts of (\ref{eq:pi}) 
we choose a simple form which can mimic the results of various calculations 
of in-medium vector mesons:
\begin{equation}
\Pi_v^{\alpha \beta }(\nu)=Z_v^{*-1}
\left((\nu -i\Gamma_v^{*}/2)^2-m_v^{*2}\right)T^{\alpha \beta},
\,\,\,\,\,\,\,   v=\rho, \omega.
\end{equation}
The asterisk denotes here the in-medium values of the resonance
position, $m_v$, width, $\Gamma_v$, and the wave-function
renormalization, $Z_v$. In the case ${\bf q}=0$ the tensor $T^{\alpha
\beta }$ has a simple form, $T^{\alpha \beta }={\rm
diag}(0,1,1,1)$ \cite{jean,chin}.  Our parameterization incorporates
basic features of mesons propagating in nuclear medium, such as the
shift of the resonance position, broadening, and wave-function
renormalization.

Applying the same formalism \cite{chin} as in the previous Section,
we find that the off-diagonal matrix element in Eq. (\ref{eq:pi}),
describing the mixing of $\rho^0$ and $\omega$ (at ${\bf q}=0$) is
given by expression
\begin{eqnarray}
\Pi_{\rho \omega}^{\alpha \beta}(\nu,\mbox{\boldmath $q$}=0) & = &
-i \int \frac{d^4k}{(2\pi)^4} \,
\left \{ {\rm Tr}[V_{\rho}^{\alpha}(\nu) G^p_D(k^0+\nu,{\bf k})
  V_{\omega}^{\beta}(-\nu) G^p_F(k)] - \right .  \nonumber \\
& & \left . {\rm Tr}[V_{\rho}^{\alpha}(\nu) G^n_D(k^0+\nu,{\bf k})
  V_{\omega}^{\beta}(-\nu) G^n_F(k)] \right \}
+ (F \leftrightarrow D) \nonumber \\
& & \equiv T^{\alpha \beta} \Pi_{\rho \omega}(\nu) ,
\label{eq:pol}
\end{eqnarray}
where $G^{p,n}_D(k)$ and $G^{p,n}_F(k)$ denote the density part and
the free part of the Dirac propagator for the proton and neutron
(compare our notation in Eq. (\ref{Nprop})). The quantity
\begin{equation}
V_{v, \alpha}  =  g_v \left( \gamma_\alpha -
\frac{\kappa_v}{2M} \sigma _{\alpha \beta } \partial^\beta \right)
\label{lint}
\end{equation}
is the vector-mesons nucleon vertex which includes the tensor coupling
$\kappa$. Following Ref.~\cite{hatsuda} we use two parameter sets:
\begin{eqnarray}
{\rm I:} & & \;\;\; g_\rho=2.63, \; \kappa_\rho=6.0, \;  g_\omega=10.1, \;
\kappa_\omega=0.12, \nonumber \\
{\rm II:} & & \;\;\; g_\rho=2.72, \; \kappa_\rho=3.7, \;  g_\omega=10.1, \;
\kappa_\omega=0.12. \nonumber
\end{eqnarray}
This parameterization follows from the vector meson dominance model
\cite{vecdom}.  The basic difference between the two sets is the value
of $\kappa_\rho$ \cite{kapparho}.

Explicit evaluation gives
\begin{eqnarray}
&& \Pi_{\rho \omega}(\nu) = \frac{2}{3} g_\rho
g_\omega \int \frac{d^3k}{(2 \pi)^3 E^*_k}
\frac{\theta(k_n-\mid {\bf k} \mid)-\theta(k_p - \mid {\bf k} \mid) }
{\nu^2-4(E^*_k)^2 } \times  \nonumber \\
&& \left [ 8 (E^*_k)^2+4 M^{*2} +
3 (\kappa_\rho+\kappa_\omega) \frac{M^*}{M} \nu^2 +
\kappa_\rho \kappa_\omega \frac{(E^*_k)^2+2 M^{*2}}{M^2} \nu^2 \right ] ,
\label{eq:polexp}
\end{eqnarray}
where $k_p$ and $k_n$ are the proton and neutron Fermi momenta and
$M^*$ is the nucleon mass in medium.  In symmetric matter, where $k_p
= k_n = k_F$, the proton and neutron contribution to
Eqs.~({\ref{eq:pol}) and (\ref{eq:polexp}) cancel, and $\Pi_{\rho
\omega}(\nu)$ vanishes. In asymmetric matter $k_n > k_p$, and we get a
net contribution to $\Pi_{\rho \omega}(\nu)$.
We note that the proton and neutron densities are equal to
\mbox{$\rho_{p,n}=k_{p,n}^3/(3 \pi^2)$}, and the baryon density
$\rho_B$ and the isospin asymmetry $x$ are equal to
\mbox{$\rho_B=\rho_{p}+\rho_{n}$} and 
\mbox{$x = (\rho_{n}-\rho_{p})/\rho_{B}$}.
At low $x$ it can be easily shown that $\Pi_{\rho \omega}(\nu)$ is
linear in $x$.  It remains linear for asymmetries accessible in
heavy-ion collisions.  If in addition we expand Eq.~(\ref{eq:polexp})
at small $\rho_B$, we notice that
\mbox{$\Pi_{\rho \omega}(\nu) \sim x \rho_B = \rho_n - \rho_p$}, in agreement
with the low-density theorem for the scattering amplitude.

Finding the eigenvalues of the matrix (\ref{eq:pi}) is
equivalent to solving the following equation
\begin{equation}
{\rm Det} \left(
\begin{array}{cc}
(\nu-i \Gamma_\rho^*/2)^2-m_\rho^{*2} &
\sqrt{Z_\rho^* Z_\omega^*}\Pi_{\rho \omega}(\nu) \\
\sqrt{Z_\rho^* Z_\omega^*}\Pi_{\rho \omega}(\nu) &
(\nu-i \Gamma_\omega^*/2)^2-m_\omega^{*2}
\end{array} \right) = 0.
\label{eq:det}
\end{equation}
Equation (\ref{eq:det}) yields eigenvalues $\nu_1$ and $\nu_2$,
and the corresponding eigenstates
$| 1 \rangle$ and $| 2 \rangle$. Our convention is that
in the absence of mixing,
{\em i.e.} for $x=0$, we have \mbox{$|1 \rangle = | \rho \rangle$} and
\mbox{$| 2 \rangle = | \omega \rangle$}.
A commonly used measure of mixing of states is the mixing angle.
Since the problem (\ref{eq:det}) is not hermitian, the eigenstates
$|1 \rangle$ and $|2 \rangle$
are not orthogonal and we cannot define a single mixing
angle. We find it useful to
introduce two mixing angles, $\theta_1$ and $\theta_2$, through the relations
\begin{equation}
| 1 \rangle = \cos \theta_1 | \rho \rangle +
\sin \theta_1 | \omega \rangle, \;\;\;
| 2 \rangle = -\sin \theta_2 | \rho \rangle +
\cos \theta_2 | \omega \rangle .
\label{eq:angles}
\end{equation}
Since the matrix in (\ref{eq:det}) is complex, the mixing angles are
also complex.

Our results are shown in Table I, which contains 8 representative
cases for $\rho_B=2 \rho_0$.  We assume that at this density
$M^*/M=0.5$.  The table should be read from top to bottom. The first
row labels the case.  Five input rows contain $m^*_\rho$,
$m^*_\omega$, $\Gamma^*_\rho$, $\Gamma^*_\omega$ and $\sqrt{Z^*_\rho
Z^*_\omega}$ for symmetric matter of density $\rho_B$.
\begin{table}[tbh]
\begin{center}
\begin{tabular}{|l|c|c|c|c|c|c|c|c|c|c|c|c|}
\hline
 Input:                       & 1 & 2 & 3 & 4 & 5 & 6 & 7 & 8   \\
\hline
$m^*_\rho$ (MeV)              &500&500&500 &550&550&650&550&550 \\
$m^*_\omega$ (MeV)            &500&500&500 &450&450&450&450&450 \\
$\Gamma^*_\rho$ (MeV)         & 0 &200&200 &300&300&300&300&300 \\
$\Gamma^*_\omega$ (MeV)       & 0 & 50& 50 & 50&200&200& 50&200 \\
$\sqrt{Z^*_\rho Z^*_\omega}$  &0.7&0.7&0.35&0.7&0.7&0.7&0.7&0.7 \\
\hline
\multicolumn{1}{|c|}{Output:} & \multicolumn{6}{|c|}{Set I}
 & \multicolumn{2}{|c|}{Set II} \\
\hline
${\rm Re}(\nu_1)$  (MeV)           &535&509&502 &557&559&656&555&557 \\
${\rm Re}(\nu_2)$ (MeV)            &469&494&499 &445&443&446&446&444 \\
$2{\rm Im}(\nu_1)$ (MeV)           & 0 &170&193 &298&306&309&297&302 \\
$2{\rm Im}(\nu_2)$ (MeV)           & 0 & 83& 57 & 56&198&198& 55&200 \\
${\rm Re}(\theta_1)$ (deg)         &45 & 12&  3 & 11& 17& 11&  9& 14 \\
${\rm Re}(\theta_2)$  (deg)        &45 & 11&  2 &  8& 13&  7&  7& 12 \\
${\rm Im}(\theta_1)$  (deg)        & 0 &-31&-13 & -7& -1&  2& -6& -2 \\
${\rm Im}(\theta_2)$  (deg)        & 0 &-32&-13 & -6& -2&  0& -6& -2 \\
\hline
\end{tabular}
\end{center}
\caption{$\rho$-$\omega$ mixing in asymmetric matter, $x=x_{\rm Pb}$,
$\rho_B=2\rho_0$, $M^*/M=0.5$.}
\label{tab:results}
\end{table}

To summarize, we observe that the mixing effects are sizable for all 
sensible cases, with mixing angles of the order of $10^o$, or larger. 
As a consequence, the resonance positions and widths of the vector
mesons are shifted significantly. Our analysis shows also (see \cite{rhoom}
for more details) that the mixing effect will continue to be important at
moderate temperatures. Therefore we expect that our results of large
$\rho$-$\omega$ mixing may show up, among other possible
medium-induced effects, in future high-accuracy relativistic heavy-ion
experiments. In particular, the results to be obtained with the HADES
spectrometer at the SIS accelerator at GSI, whose anticipated mass
resolution in the discussed region will reach 1\% \cite{hades},
should be influenced by the phenomenon of $\rho$-$\omega$ mixing.

\end{document}